\documentclass[aps,showpacs,prl,twocolumn]{revtex4-1}  % for review and submission 

\usepackage{amsmath}
\usepackage{mathrsfs}
\usepackage{amssymb}
\usepackage{graphicx,epstopdf,epsfig}
\usepackage{bm}
\usepackage{color}
\usepackage{times}
\usepackage{hyperref}

\newcommand{\mr}[1]{{\mathrm{#1}}} 		% mathrm shortcut
\newcommand{\bra}[1]{{\langle #1 \vert}} 		%bra < x |
\newcommand{\ket}[1]{{\vert #1 \rangle}} 		%ket | x >

\newcommand{\ii}{\,\mr{i}}					%Complex i
\definecolor{clr}{rgb}{0,0.6,0.6}

\begin{document}

\title{Hybrid Atom--Photon Quantum Gate in a Superconducting Microwave Resonator}
\author{J. D. Pritchard}
\author{J. A. Isaacs}
\author{M. A. Beck}
\author{R. McDermott}
\author{M. Saffman}
\email{msaffman@wisc.edu}
\affiliation{Department of Physics, University of Wisconsin, 1150 University Avenue, Madison, Wisconsin 53706}

\date{\today}

\begin{abstract}
We propose a novel hybrid quantum gate between an atom and a microwave photon in a superconducting coplanar waveguide cavity by exploiting the strong resonant microwave coupling between adjacent Rydberg states. Using experimentally achievable parameters gate fidelities $> 0.99$ are possible on sub-$\mu$s timescales for waveguide temperatures below 40~mK. This provides a mechanism for generating entanglement between two disparate quantum systems and represents an important step in the creation of a hybrid quantum interface applicable for both quantum simulation and quantum information processing.\end{abstract}
%Quantum Optics 42.50.Ar
%Rydberg Atoms 32.80.Rm
%Quantum Computation 03.67.Lx,
%Photon-Atom Interactions 32.80.-t
\maketitle

Quantum information processing provides a route to efficient solutions for many problems that are intractable on conventional classical computers. While there has been tremendous recent progress in the realization of small-scale quantum circuits comprised of several quantum bits (``qubits") using a wide range of physical implementations \cite{ladd10}, research indicates that a fault-tolerant quantum computer that exceeds what is possible on existing classical machines will require a network of thousands of qubits, far beyond current capabilities. 

Hybrid quantum computation exploits the unique strengths of disparate quantum technologies, enabling realization of a scalable processor capable of both fast gates and long coherence times that has yet to be achieved using a single physical qubit. Superconducting qubits coupled via microwave resonators have been identified as promising candidates for such an interface offering both fast ($\sim$ns) gate times and a scalability through fabrication of chip-based superconducting circuits \cite{wallraff04,blais04,xiang13}, which have already been used to implement quantum algorithms \cite{dicarlo09}, many-body Hamiltonians \cite{fink08} and synthesis of arbitrary quantum states of the resonator mode \cite{hofheinz09}. However, a limiting factor is the coherence time of the qubits, typically around 60~$\mu$s \cite{chang13,barends13}. This makes coupling the superconducting circuits to external qubits for quantum memory advantageous. To date a number of systems have been explored for this purpose, including solid state spin-ensembles \cite{kubo11}, color centers in diamond \cite{amsuss11}, nano-mechical resonators \cite{oconnell10} and atoms.

Atomic systems offer very long coherence times, and can be coupled to the microwave cavity mode via the weak magnetic dipole transition between hyperfine ground states \cite{verdu09,petrosyan09,hafezi12}. Recently, coherence times exceeding $3$~s have been demonstrated for a BEC held above a resonator at 4~K \cite{bernon13}. Alternatively, the strong electric dipole coupling between close-lying Rydberg states can be used to obtain strong coupling of a single atom to the microwave field\cite{sorensen04,petrosyan09,patton13}. Rydberg atoms have strong dipole-dipole interactions which can be utilised for both atomic and photonic quantum gates \cite{saffman10,pritchard13}. These combined properties can be exploited to create a hybrid quantum interface capable of fast processing times using the superconducting qubits with long storage time in the atomic quantum memory, as well as the ability to map from the atomic state to optical systems for communication across a quantum network \cite{kimble08}. The microwave resonator can additionally be utilised to extend the strong interactions between Rydberg atoms to enable entanglement of single photons interacting with spatially separated atomic ensembles \cite{petrosyan08}. Recent progress towards chip-based experiments has been the demonstration of coherent driving of Rydberg states near the surface of atom chips \cite{hogan12,carter13}. 

\begin{figure}[t]
\includegraphics{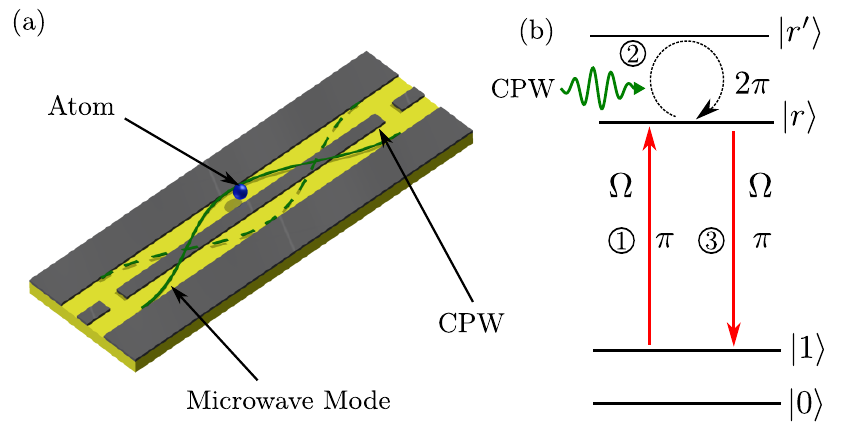}
\caption{(Color online). Hybrid Quantum Gate Scheme (a) An atom trapped above the electric field antinode of a superconducting microwave coplanar waveguide resonator. (b) Level-scheme for atom--photon conditional phase gate. The atom is prepared using hyperfine ground states as the qubit basis. A $\pi$-pulse excitation maps $\ket{1}\rightarrow\ket{r}$, which is resonantly coupled to close-lying state $\ket{r'}$ via the microwave cavity and undergoes a $2\pi$-rotation conditional upon the presence of a cavity photon. A second $\pi$-pulse from $\ket{r}\rightarrow\ket{1}$ with the resulting phase controlled by the microwave photon.}
\label{fig1}
\end{figure}

%to provide long coherence quantum memory, strong coupling atoms \cite{verdu09,petrosyan09} coherence above waveguide \cite{bernon13}

%Q in excess of 1 million \cite{megrant12}. Rydberg atom capacitively coupled to conductor \cite{sorensen04,tian04} nano-mechanical resonator\cite{gao11} ensembles for photon gates \cite{petrosyan08} . Rydberg atoms excited over microwave guide\cite{hogan12}. Rydberg atom quantum information \cite{saffman08,isenhower10}. Cavity QED modification of spontaneous emission \cite{kleppner81,hulet85} Maser \cite{meschede85} quantum non-demolition \cite{guerlin07}.

We propose a novel hybrid quantum gate between a superconducting microwave cavity photon and an atom trapped above the electric field antinode of a coplanar waveguide resonator (CPW) as shown in Fig.~\ref{fig1} which exploits the strong electric dipole moment $d\sim n^2ea_0$ between close-lying Rydberg states with transition frequencies in the microwave regime, where $n$ is the principal quantum number. The electric dipole of the $n=100$ Rydberg state is approximately 6 orders of magnitude greater than the magnetic dipole coupling of the hyperfine ground states, making it possible to achieve strong coupling with a single atom rather than requiring a $\sqrt{N}$ collective coupling enhancement \cite{verdu09,petrosyan09}. This gate can be used to transfer entanglement from the cavity mode to the atomic state, facilitating an atomic quantum memory for superconducting circuits which is a vital component for constructing a robust hybrid quantum interface. Unlike previous proposals \cite{petrosyan09} this scheme only requires a single resonant coupling from an atomic qubit level to the Rydberg state, simplifying the implementation.

We consider an atom held close to the surface of a superconducting CPW cavity with a cavity frequency $\omega_c$ and cavity lifetime $1/\kappa$. A schematic of the experiment is shown in Fig.~\ref{fig1}(a). The relevant atomic level scheme for the gate implementation is shown in Fig.~\ref{fig1}(b), with the qubit states encoded in two different hyperfine ground states of an alkali atom. State $\ket{1}$ is resonantly coupled to a high principal quantum number Rydberg state $\ket{r}$ via a classical laser field with Rabi frequency $\Omega$. This Rydberg state is chosen such that the transition frequency, $\omega_{rr'}=\omega_{r'}-\omega_{r}$, from $\ket{r}$ to a close lying ($\Delta n=0$) microwave coupled state $\ket{r'}$ is nearly resonant with the microwave cavity mode, $\omega_c$. The choice of $\Delta n=0$ maximizes the dipole moment $\bm{d}_{rr'}$ for the transition. The coupling between the Rydberg states and the cavity is described by the Jaynes-Cummings Hamiltonian \cite{jaynes63}, which in the rotating wave approximation is given by  %Different $\Delta n$ can be chosen to tune the coupling strength for other non-circuit model applications.
\begin{equation}
\hat{\mathscr{H}} = \hbar\Delta\hat{\sigma}_+\hat{\sigma}_- + \hbar g(\hat{\sigma}_+\hat{a} + \hat{\sigma}_-\hat{a}^\dagger),
\end{equation}
where $\Delta=\omega_c-\omega_{rr'}$, $\hat{\sigma}_+=\ket{r'}\bra{r}$ is the Rydberg state atomic raising operator, $\hat{a},\hat{a}^\dagger$ the annihilation and creation operators for the microwave cavity field and $g=\bm{\mathcal{E}}_{0}(\bm{x}_a)\cdot\bm{d}_{rr'}/\hbar$ is the vacuum coupling strength between an atom and the cavity which is determined by the zero-point electric field of the cavity mode at the position of the atom, $\bm{x}_a$.

In the strong coupling regime with $g\gg\kappa,\gamma_{r,r'}$ where $1/\gamma_{r,r'}$ is the lifetime of the Rydberg states, when the atom is on resonance with the cavity ($\Delta=0$) then a single photon in the cavity will drive coherent vacuum Rabi oscillations between $\ket{r}$ and $\ket{r'}$ at a frequency of $2g$. This can be utilized to realize a conditional phase gate via the following scheme: 1) The cavity is initially detuned from the atomic resonance with $\Delta\gg g,\kappa,\gamma$ to minimize the perturbation of the atomic energy levels due to the coherent coupling to the cavity energy levels. The classical laser field is turned on for a time $\tau_1=\pi/\Omega$ to drive a high fidelity $\pi$ pulse from the upper qubit state to the Rydberg state $\ket{1}\rightarrow \ii\ket{r}$. 2) The cavity is tuned into resonance with $\Delta=0$ for a period of $\tau_2=\pi/g$. If a photon is present in the cavity it drives a $2\pi$ rotation of the Rydberg state $\ii\ket{r}\rightarrow-\ket{r'}\rightarrow-\ii\ket{r}$, otherwise the Rydberg state is unchanged. 3) The cavity is detuned again and an identical $\pi$-pulse as in 1) is applied to map the Rydberg state back to the upper qubit state $\pm\ii\ket{r}\rightarrow\mp\ket{1}$ in a duration $\tau_3=\tau_1$ conditional upon the photon occupation number of the cavity mode. This thus realizes a $C_z$ conditional phase gate with $C_z=\ket{00}\bra{00}-\ket{01}\bra{01}+\ket{10}\bra{10}+\ket{11}\bra{11}$, where the two qubit basis is defined as first the CPW Fock state and second the atomic basis state. The total duration for the gate operation is given by $\tau=\pi(2/\Omega+1/g)$. Combining this universal gate with single qubit rotations it is therefore possible to perform arbitrary two-qubit quantum gates and provides a route to entangle the atom-photon system. %any quantum algorithm on the hybrid atom--cavity system, for example performing a Hadamard gate $H$ before and after the $C_z$ creates a CNOT gate.

This simple gate sequence relies on the ability to rapidly change $\Delta$ on a timescale fast compared to $g$, which can be achieved either using a Stark-shift of the Rydberg states with an external DC electric field to change $\omega_{rr'}$ or by using an intra-cavity Superconducting Quantum Interference Device (SQUID) which can be used to switch $\omega_c$ by 0.5~GHz in 1~ns \cite{sandberg08,wang13}. 

{\color{black}{To evaluate the gate performance and demonstrate entanglement generation we calculate the preparation fidelity of the Bell state $\ket{\Psi^+}=\frac{1}{\sqrt{2}}(\ket{01}+\ket{10})$ with the atom initially in $\ket{0}$ and the resonator in a superposition state $(\ket{0}+\ket{1})/\sqrt{2}$. Entanglement is established using the gate sequence $H_aC_zH_a$, where $H_{a}$ denotes a Hadamard gate applied to the atom. Initialisation of the resonator is achieved by controllably mapping the state of a superconducting qubit onto the cavity mode by shifting the qubit into resonance for a time $\pi/2g_{SC}$ \cite{mariantoni11}. This preparation step is included in the simulation using typical properties of a superconducting qubit with $g_{SC}/2\pi=100~$MHz and a lifetime of 2~$\mu$s \cite{blais04}, however we assume perfect qubit initialisation and single qubit operations to determine the intrinsic hybrid gate fidelity.}}

We explicitly consider the cavity coupling from the 90$s_{1/2}, m_j=1/2$ to 90$p_{3/2}, m_j=1/2$ in Cs with transition frequency $\omega_{rr'}=2\pi\times5.037$~GHz and a dipole moment of $d_{rr'}=\sqrt{2/9}\times8360~ea_0$. The radiative lifetimes of the states are 820~$\mu$s and 2~ms respectively \cite{beterov09}, with negligible contribution from the black-body radiation at the milli-Kelvin temperatures used for performing superconducting circuit experiments. Similarly, the $A$ coefficient for the decay from 90$p_{3/2}$ to 90$s_{1/2}$ is 0.2 $s^{-1}$ and can also be neglected. The time evolution of the density matrix $\rho$ is calculated using the Lindblad master equation

%\begin{equation}
%\begin{aligned}
%\dot{\rho}=&-\frac{i}{\hbar}[\mathscr{H},\rho]+\frac{\kappa}{2}(\bar{n}_\mr{th}+1)(2\hat{a}\rho\hat{a}^\dagger-\hat{a}^\dagger\hat{a}\rho-\rho\hat{a}^\dagger\hat{a})\\
%&+\frac{\kappa}{2}\bar{n}_\mr{th}(2\hat{a}^\dagger\rho\hat{a}-\hat{a}\hat{a}^\dagger\rho-\rho\hat{a}\hat{a}^\dagger)\\
%&+\frac{\gamma_r}{2}(2\hat{\sigma}_{rs}^-\rho\hat{\sigma}_{rs}^+-\hat{\sigma}^+_{rs}\hat{\sigma}^-_{rs}\rho-\rho\hat{\sigma}^+_{rs}\hat{\sigma}_{rs}^-)\\
%&+\frac{\gamma_{r'}}{2}(2\hat{\sigma}_{r's}^-\rho\hat{\sigma}_{r's}^+-\hat{\sigma}^+_{r's}\hat{\sigma}^-_{r's}\rho-\rho\hat{\sigma}^+_{r's}\hat{\sigma}_{r's}^-),
%\end{aligned}
%\end{equation}

\begin{figure}[t]
\includegraphics[width=8.6cm]{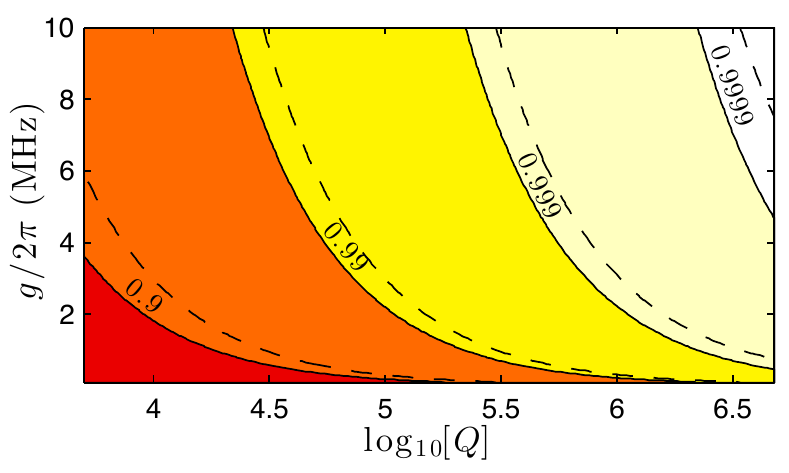}
\caption{(Color online). Contour plot of the $\ket{\Psi^+}$ Bell state preparation fidelity $F$ using the hybrid quantum gate. At high $Q$ the fidelity is limited by spontaneous decay of the Rydberg levels during the conditional rotation time $\tau=\pi/g$. Dashed lines show analytic fidelity estimate.}
\label{fig2}
\end{figure}

\begin{equation}
\dot{\rho}=-\frac{i}{\hbar}[\mathscr{H},\rho]+\frac{1}{2}\displaystyle\sum_i(2\hat{c}_i\rho\hat{c}_i^\dagger-\hat{c}^\dagger_i\hat{c}_i\rho-\rho\hat{c}^\dagger_i\hat{c}_i),
\end{equation}
where the decay operators $\hat{c}_i$ are given by
\begin{subequations}
\begin{align}
\hat{c}_1&=\sqrt{\gamma_r}\hat{\sigma}_{rs}^-,\\
\hat{c}_2&=\sqrt{\gamma_{r'}}\hat{\sigma}_{r's}^-,\\
\hat{c}_3&=\sqrt{\kappa(\bar{n}_\mr{th}+1)}\hat{a},\\
\hat{c}_4&=\sqrt{\kappa\bar{n}_\mr{th}}\hat{a}^\dagger,
\end{align}
\end{subequations}
where it is assumed the population decaying from the Rydberg states ends up in alternative ground states to give a lower limit on the achievable fidelity due to spontaneous emission as in practice a fraction of the population will decay back to the qubit states. This is modelled by inclusion of an additional reservoir state $\ket{s}$ coupled via raising and lowering operators $\hat{\sigma}_{r,r's}^\pm$ respectively. The photon decay rate from the cavity is determined by $\kappa$ which is related to the quality factor of the resonator by $Q=\omega_c/\kappa$, in addition to a contribution arising from the mean thermal photon number within the cavity $\bar{n}_\mr{th}$. Finite temperature effects will be considered below but initial calculations are performed in the zero-temperature limit with $\bar{n}_\mr{th}=0$. The evolution is calculated for a range of coupling strengths $g$ and $Q$-factors, with the fidelity determined using $F=\mr{Tr}(\sqrt{\sqrt{\rho_i}\rho\sqrt{\rho_i}})$, where $\rho_i=\ket{\Psi^+}\bra{\Psi^+}$ is the ideal density matrix for the Bell state.

The results are plotted in Fig.~\ref{fig2} which shows a contour map of the Bell state preparation fidelity $F$ for a range of $g$ and $Q$, showing high fidelity gates with $F>0.999$ are achievable. At low $Q$, the preparation fidelity is limited by the short lifetime of the photons within the cavity mode. As $Q$ is increased the finite lifetime of the Rydberg states dominates, leading to a requirement for higher $g$ to reduce the time needed for the conditional rotation with the cavity mode. {{\color{black}{Taking account of the gate errors caused by decay of the atom or photon during the entanglement preparation results in an analytic fidelity estimate of $F=1-\pi/(16g)*[3(\kappa+\gamma_r)+\gamma_{r'}]$, which is plotted as dashed contours on Fig.~\ref{fig2} and shows reasonable agreement with the complete numerical simulation. For the parameters considered, $\kappa$ is the dominant source of error requiring a $Q>2\times10^7$ to be comparable with the finite lifetimes of the $n=90$ Rydberg states which has yet to be demonstrated using surface fabricated waveguides.}}

The simulation neglects errors accumulated during the $\pi$-pulses between $\ket{1}$ and $\ket{r}$, during which time the dominant error arises from spontaneous emission from the Rydberg state which occurs with a probability $P_r=2(\pi/\Omega)(\gamma_r/2)$, which for $\Omega/2\pi=10$~MHz gives $P_r\sim6\times10^{-5}$ which is small compared to a fidelity of $F>0.999$. It is interesting to note that the potential state fidelity of the hybrid gate is nominally higher than for a two-atom quantum gate based on the Rydberg blockade mechanism \cite{zhang12a}. This is partly due to the cryogenic environment increasing the Rydberg state lifetime, and partly because errors arising from the imperfect atom-atom blockade interaction are avoided. However, a complete quantum process tomography is required to evaluate the full gate fidelity.

We now consider the requirements for practical implementation of such a gate, starting with realistic values for $g$ and $Q$. The workhorse of superconducting circuit QED is the microwave CPW \cite{blais04} in Fig.~\ref{fig1}(a) which features a central conductor between two large ground planes which is fabricated on a dielectric substrate which supports quasi-TEM microwave fields. The resonator frequency is determined geometrically and can be modified through choice of length and the ratio of the widths of the center conductor, $s$, to the gap to the ground plane, $w$ \cite{goppl08}. The zero-point energy of the cavity field $\hbar\omega_c/2$ is shared equally between the electric and magnetic field modes of the CPW making it possible to normalize the zero-point electric field in the cavity calculated using finite element analysis via the energy density using
\begin{equation}
\frac{\hbar\omega_c}{2}=\displaystyle\int\epsilon(r)\vert \bm{\mathcal{E}_0}(\bm{r})\vert ^2\mr{d}^3\bm{r},
\label{eq2}\end{equation}
where $\epsilon(\bm{r})$ is the permittivity equal to $\epsilon_0$ above the cavity and $\epsilon_r\epsilon_0$ in the dielectric substrate. Figure~\ref{fig3} shows the transverse zero-point electric field magnitude and direction at the anti-node of a 5~GHz superconducting $\lambda/2$ resonator grown on sapphire ($\epsilon_r=9.6$) with a center trace width of $s=20~\mu$m and gap of $w=10~\mu$m, typical of CPW used for superconducting qubits \cite{blais04}. The peak field is confined within the gap between traces at $x=(s+w)/2$ and falls off approximately exponentially above the surface, as shown in Fig.~\ref{fig3}(b) which also plots the coupling strength $g$ calculated using the dipole moment of $d_{rr'}$ given above, resulting in a peak coupling of $g/2\pi=4$~MHz. 

\begin{figure}[t]
\includegraphics[width=8.6cm]{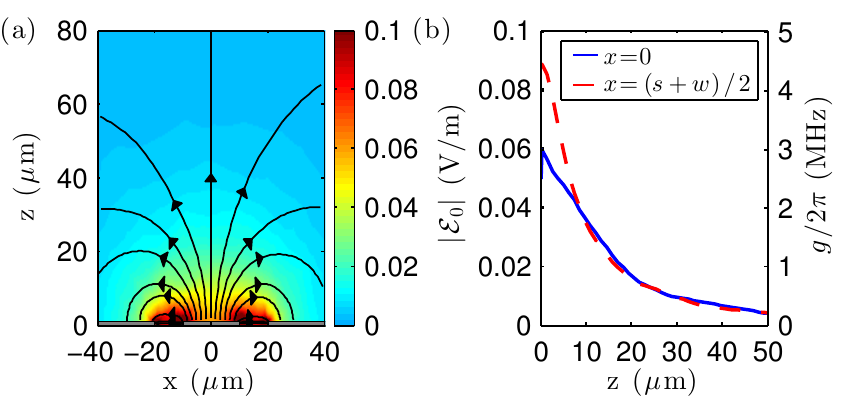}
\caption{(Color online). (a) Zero-point electric field distribution around superconducting CPW cavity with a center trace of 20~$\mu$m and a 10$~\mu$m gap. (b) Field distribution and vacuum coupling strength to the 90$s_{1/2}\rightarrow90p_{3/2}$ transition as a function of height above the surface for both the resonator midpoint and gap centre, which falls off rapidly on a length scale comparable to the slot width $w$}.
\label{fig3}
\end{figure}

Trapping atoms close to superconducting circuits presents a significant challenge, as light scattered onto the superconductor can create localized heating and quasi-particle excitations that adversely affect the cavity $Q$. Currently efforts have focused on the magnetic trapping of atoms above superconducting waveguides, making it possible to achieve $z=10~\mu$m \cite{bernon13} corresponding to $g/2\pi\sim2~$MHz. Whilst increasing the quantum number of the Rydberg states will increase the dipole moment by $n^2$ this concomitantly reduces the cavity frequency proportional to $n^{-3}$ and hence the zero point electric field \footnote{Inspection of eq.~\ref{eq2} implies $\mathcal{E}_0\propto\sqrt{\omega_c}$, however the change in the resonator length to maintain the same resonant mode introduces an additional factor of $\sqrt{\omega_c}$.} resulting in an effective reduction of $g\propto 1/n$. An important figure of merit for the cavity QED system is however the ratio of $g/\gamma_{r}\propto n^2$, making $n=90$ an ideal compromise between $g$ and the Rydberg state lifetime $\propto n^3$ whilst operating at a frequency matched to state-of-the-art superconducting quantum circuits \cite{blais04,goppl08}. {\color{black}{An additional difficulty associated with trapping Rydberg states close to a surface arises from the large dc polarisability $\propto n^7$ resulting in large Stark shifts and dephasing from electric field fluctuations. Patch potentials due to atomic adsorbates have been observed to create a significant contribution to the background field which increases over time \cite{tauschinsky10,hattermann12}, so using a single atom will significantly reduce the deposition rate. Similarly, electric field fluctuations close to a gold surface at 300~K have been studied using Rydberg superposition states \cite{carter13} however at cryogenic temperatures there is significant reduction of the field noise \cite{labaziewicz08} corresponding to a dephasing rate of order 10~Hz at $n=90$. This effect can be further suppressed by using off-resonant dressing fields to null the differential ac Stark shift at low frequencies for states $\ket{r}$ and $\ket{r'}$. \cite{jones13}.}}

The second critical cavity parameter is the intrinsic quality factor of the resonator, $Q$ which determines the photon lifetime within the cavity. This is predominantly limited by the losses inherent in the dielectric medium and the presence of contaminant particles on the chip surface. The best resonators to date have a $Q>10^6$ \cite{megrant12}; however typically $Q\sim10^5$ \cite{goppl08}. Using these parameters mean state fidelities of $F>0.99$ can be achieved, as shown in Fig.~\ref{fig2}. Assuming a classical coupling rate of $\Omega/2\pi=10$~MHz, the total gate duration is 350~ns.

Finally, we consider the effects of finite temperature on the gate fidelity. Unlike optical cavity systems with frequencies in the THz regime, superconducting microwave cavities have significant thermal occupation even at cryogenic temperatures. The mean thermal photon number in the cavity is given by $\bar{n}_\mr{th}=(e^{\hbar\omega_c/k_BT}-1)^{-1}$, which causes incoherent driving of the $\ket{r}\rightarrow\ket{r'}$ transition in addition to reducing the fidelity of the cavity photon state preparation. To demonstrate the dependence on CPW temperature we calculate the Bell state preparation fidelity as a function of temperature for the gate detailed above using realistic parameters of $g/2\pi=2$~MHz and $Q=2\times10^5$. 

{\color{black}{The results are plotted in Fig.~\ref{fig4}, which shows both the fidelity of the cavity preparation in state $(\ket{0}+\ket{1})/\sqrt{2}$ following interaction with the superconducting qubit, $F_\gamma$, and the Bell state preparation fidelity $F_{\Psi^+}$ for the entire gate sequence. This shows that whilst the gate evolution is insensitive at low  temperature the fidelity drops rapidly above 40~mK due to the thermal photon occupation preventing high fidelity preparation of the cavity mode. This represents a physical limitation for any experiments performed at this frequency, necessitating use of a dilution refrigerator. Local heating of the resonator due to the excitation lasers can be minimised using a two-photon transition from $\ket{g}$ to $\ket{r}$ with a large single-photon detuning relative to the intermediate state to provide stronger transition matrix elements than the single photon transition, reducing the required optical intensity at the atoms. Additionally, the single atom may be replaced by an ensemble and the Rydberg dipole blockade mechanism used to excite a single collective excitation \cite{lukin01}, giving a $\sqrt{N}$ enhancement in the coupling strengths.}}

\begin{figure}[t]
	\includegraphics[width=8.6cm]{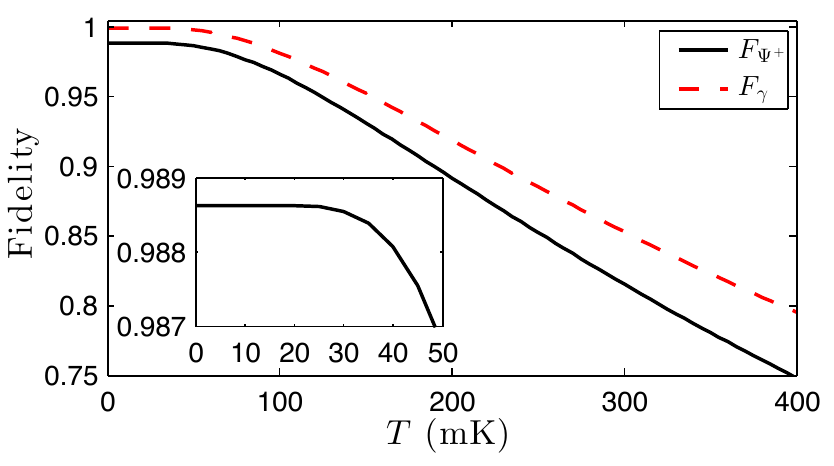}
	\caption{(Color online) Finite temperature effects due to thermal occupation of the cavity mode. $F_\gamma$ represents the fidelity of preparing the cavity in state $(\ket{0}+\ket{1})/\sqrt{2}$ via resonant interaction with a superconducting qubit which drops rapidly above 40~mK due to the thermal occupation of the cavity. $F_{\Psi^+}$ is the resulting Bell-state preparation fidelity which follows approximately linearly with $F_\gamma$. Inset: zoom in around zero temperature. \label{fig4}}
\end{figure}

{\color{black}{The preceding results demonstrate that the hybrid controlled phase gate can generate entanglement between an atomic qubit and cavity field inside a microwave resonator with high fidelity. To extend the gate to application as a quantum memory for a superconducting qubit Q$_1$, a teleportation sequence \cite{bennett93} can be used requiring an ancillary superconducting qubit Q$_2$. Following preparation of the atom-photon Bell state, the cavity field is then mapped to Q$_2$. Bell state measurements are then performed on Q$_1$ and Q$_2$, with the outcome used to apply the required single qubit rotations onto the atom. This transfers an arbitrary state $\ket{\psi}$ from qubit Q$_1$ to the atomic qubit to exploit the long atomic coherence time. Bell state measurements on superconducting qubits can be performed and processed in $\sim500$~ns \cite{steffen13}, offering much faster storage compared to construction of a swap gate from the controlled phase gate above.}}

%{\color{black}{The results above demonstrate the hybrid controlled phase gate enables the generation of entanglement between an atomic qubit and cavity field inside a microwave resonator with high fidelity. To extend the gate to application as a quantum memory for a superconducting qubit, the following protocol can be used. First the atom and cavity are prepared in $\ket{0}$, then the superconducting qubit state $\ket{\psi}$ is mapped onto the resonator mode by tuning it to resonance with the cavity \cite{mariantoni11}. Next a $\mr{SWAP}$-gate between the atom and cavity is realised as three consecutive $CNOT$ gates alternating between the atom and photon as the control qubit via the gate sequence $(H_aU_{C_z}H_a)(H_\gamma U_{C_z}H_\gamma)(H_aU_{C_z}H_a)$, resulting in the atom in state $\ket{\psi}$. \textit{To perform the Hadamard gate on the cavity mode, $H_\gamma$, the off-resonant interaction with the superconducting qubit is used??}. Assuming a 1~$\mu$s duration for the atomic Hadamard gate, the time required for the SWAP gate is approximately 8~$\mu$s which is comparable to photon lifetime at $Q=10^5$.!?}}

In conclusion, we have proposed a method of entangling superconducting cavity photons with Rydberg-atom qubits which introduces few losses relative to its homogenous counterparts. Furthermore, we show that currently achievable experimental implementations in each field can produce hybrid entanglement fidelities in the $.99$ regime with gate times of order a few $\mu$s, limited only by the $Q$-factor of the resonator cavity. {{\color{black}{Utilising the coupling of the microwave resonator to a diverse range of quantum systems via the electric or magnetic dipole transition can then be exploited to map the entanglement onto an qubit such as a collective solid state spin system or a superconducting qubit fabricated within the resonator mode \cite{xiang13}. This provides the possibility to use cold atoms as quantum memories for superconducting systems, as well as offering the potential to convert microwave photons to the optical regime where qubits are protected from thermal noise.}}

\begin{acknowledgments}
This work was supported by funding from the NSF award PHY-1212448 and the University of Wisconsin Graduate School.
\end{acknowledgments}

%\bibliography{../../Papers/Hybrid}

%merlin.mbs apsrev4-1.bst 2010-07-25 4.21a (PWD, AO, DPC) hacked
%Control: key (0)
%Control: author (8) initials jnrlst
%Control: editor formatted (1) identically to author
%Control: production of article title (-1) disabled
%Control: page (0) single
%Control: year (1) truncated
%Control: production of eprint (0) enabled
%

\end{document}